\documentclass[usenatbib]{mnras}
\usepackage{lscape}
\usepackage{amsmath,amssymb}
\usepackage{mathtools}
\usepackage{mathrsfs}	
\usepackage{multicol}
\usepackage{textcomp}
\usepackage{graphicx}
\usepackage{xcolor}
\usepackage{caption}
\usepackage{subfig}
\usepackage{mdwlist}
\usepackage{ulem}

\graphicspath{{figures/}}

\catcode`\@=11
\def\gta{\ifmmode{\,\mathrel{\mathpalette\@versim>\,}}
    \else{$\,\mathrel{\mathpalette\@versim>}\,$}\fi}
\def\lta{\ifmmode{\,\mathrel{\mathpalette\@versim<\,}}
    \else{$\,\mathrel{\mathpalette\@versim<}\,$}\fi}
\def\@versim#1#2{\lower 2.9truept \vbox{\baselineskip 0pt \lineskip
    0.5truept \ialign{$\m@th#1\hfil##\hfil$\crcr#2\crcr\sim\crcr}}}
\catcode`\@=12  

{\newif\ifnotend
\notendtrue
\def\veclist{ABCDEFGHIJKLMNOPQRSTUVWXYZabcdefghijklmnopqrstuvwxyz.}
\def\top#1#2.{#1}
\def\tail#1#2.{#2.}
\loop\expandafter\xdef\csname bb\expandafter\top\veclist\endcsname%
{{\noexpand\bf\expandafter\top\veclist}}
\edef\veclist{\expandafter\tail\veclist}
\if\veclist.\notendfalse\fi\ifnotend\repeat}

\newcommand\Sigmas{\Sigma_*}
\newcommand\Ms{M_*}
\newcommand\as{a_*}
\newcommand\sigphi{\sigma_{\varphi}}

\newcommand\vphiqm{\overline{v_{\varphi}^2}}
\newcommand\sigR{\sigma_R}
\newcommand\sigz{\sigma_z}
\newcommand\Deltac{\Delta_R}

\newcommand{\phih}{\Phi_{\mathrm{h}}}
\newcommand{\rhoh}{\rho_{\mathrm{h}}}
\newcommand{\rhos}{\rho_*}

\newcommand{\bM}{b_{\mathrm{M}}}
\newcommand{\bm}{b_{\mathrm{m}}}
\newcommand{\bz}{b_0}
\newcommand{\bu}{b_1}
\newcommand{\bd}{b_2}
\newcommand{\vh}{v_{\mathrm{h}}}
\newcommand{\bplus}{\mathscr{B}^{+}}
\newcommand{\bminus}{\mathscr{B}^{-}}
\newcommand{\cplus}{\mathscr{C}^+}
\newcommand{\cminus}{\mathscr{C}^-}
\newcommand{\dplus}{\mathscr{D}^{+}}
\newcommand{\dminus}{\mathscr{D}^{-}}
\newcommand{\bpm}{\mathscr{B}^{\pm}}
\newcommand{\cpm}{\mathscr{C}^{\pm}}
\newcommand{\dpm}{\mathscr{D}^{\pm}}
\newcommand{\bzero}{\mathscr{B}^0}
\newcommand{\czero}{\mathscr{C}^0}
\newcommand{\dzero}{\mathscr{D}^0}

\newcommand{\bzplus}{\mathscr{B}_z^{+}}
\newcommand{\bzminus}{\mathscr{B}_z^{-}}
\newcommand{\czplus}{\mathscr{C}_z^+}
\newcommand{\czminus}{\mathscr{C}_z^-}

\newcommand{\dzminus}{\mathscr{D}_z^{-}}

\newcommand{\bzpm}{\mathscr{B}_z^{\pm}}

\newcommand{\projz}{\mathrm{Pr}}
\newcommand{\vphim}{\overline{v_{\varphi}}}

\newcommand{\reff}{R_{\mathrm{e}}}

\newcommand{\sB}{s_{\mathscr{B}}}
\newcommand{\sC}{s_{\mathscr{C}}}

\newcommand\MR{\mathscr{R}}

\newcommand\cu{c_1}
\newcommand\cd{c_2}
\newcommand\Comm{\left[\rho_*,\Phi\right]}
\newcommand\Commh{\left[\rho_*,\Phi_{\rm h}\right]}
\newcommand\Bx{\mathrm{B}_x}

\newcommand\gammap{\gamma_{\varphi}}

\begin{document}

\setlength{\parskip}{0pt}

\date{Accepted 2025 October 29. Received 2025 October 18; in original form 2025 July 25}
\title[Anisotropy ansatz for oblate ETGs]
{Anisotropy ansatz for the Jeans equations: oblate galaxies}
{}
\author[L. De Deo, L. Ciotti \& S. Pellegrini]{Leonardo De
  Deo$^{1,2}$, Luca Ciotti$^1$ \& Silvia Pellegrini$^{1,2}$
  \\
$^{1}$Department of Physics and Astronomy, University of Bologna, Via Gobetti 93/2, 40129 Bologna, Italy\\
$^{2}$INAF - Osservatorio di Astrofisica e Scienza dello Spazio di Bologna, Via Gobetti 93/3, 40129 Bologna, Italy
}
\maketitle
\begin{abstract}
In the solution of the Jeans equations for axisymmetric galaxy models the ``$b$-ansatz" is often adopted to prescribe the relation between the vertical and radial components of the velocity dispersion tensor, and close the equations. However, $b$ affects the resulting azimuthal velocity fields quite indirectly, so that the analysis of the model kinematics is usually performed after numerically solving the Jeans equations, a time consuming approach. In a previous work we presented a general method to determine the main properties of the kinematical fields resulting in the $b$-ansatz framework before solving the Jeans equations; results were illustrated by means of disk galaxy models. In this paper we focus more specifically on realistic ellipsoidal galaxy models. It is found that how and where $b$ affects the galaxy kinematical fields is mainly dependent on the flattening of the stellar density distribution, moderately on the presence of a Dark Matter halo, and much less on the specific galaxy density profile. The main trends revealed by the numerical exploration, in particular the fact that more flattened systems can support larger $b$-anisotropy, are explained with the aid of simple ellipsoidal galaxy models, for which most of the analysis can be conducted analytically. The obtained results can be adopted as guidelines for model building and in the interpretation of observational data.
\end{abstract}

\section{Introduction}

The Jeans equations (hereafter JEs) are widely used for the modeling of stellar systems, both in theoretical and observational studies (e.g. \citealt{Binney2008gady.book.....B}, hereafter BT08; \citealt{2014dyga.book.....B}; \citealt{Ciotti2021isd..book.....C}, hereafter C21). In the case of axisymmetric systems, the simplest and most common choice is to use the JEs obtained under the assumption that the underlying phase-space distribution function depends on the stars orbital energy and angular momentum component along the symmetry axis\footnote{From now on we will use standard cylindrical coordinates $(R,\varphi,z)$, where $z$ is the symmetry axis.}. A well known consequence of this choice is the resulting identity of the vertical $(\sigz)$ and radial $(\sigR)$ components of the velocity dispersion tensor, and the alignment at each point of the system of the velocity dispersion ellipsoid with the local basis of cylindrical coordinates\footnote{Three-integrals axisymmetric systems with a spherically aligned velocity dispersion ellipsoid have been also considered (e.g., see \citealt{Cappellari2020MNRAS.494.4819C,2024ApJ...970....1V}).}. Data however often seem to require that the two components are not equal, and so in the JEs the introduction of a relationship between $\sigz$ and $\sigR$ as a phenomenological, observationally motivated closure, is favored. A widely used choice is the $b$-ansatz (\citealt{Cappellari2008MNRAS.390...71C}), for which $\sigR^2=b\,\sigz^2$, with a suitably assigned positive value of $b$. The solution of the resulting JEs proved to capture the main properties of the stellar orbits distribution inferred from extensive three-integrals Schwarzschild modeling of integral-field stellar kinematics, under the mass-follows-light hypothesis, and to model adequately the observations, in particular those of axisymmetric galaxies classified as regular rotators and stellar disks (see e.g. \citealt{Cappellari2007MNRAS.379..418C}; see also \citealt{Cappellari2016ARA&A..54..597C} for a review). Recently, also data relative to dwarf spheroidal galaxies have been successfully modeled with the $b$-ansatz (\citealt{2020ApJ...904...45H}). 

The $b$-ansatz however, albeit phenomenologically satisfactory, affects the solution of the JEs in a non-trivial way; in particular the azimuthal velocity field $\vphiqm$, for some choices of the $b$ value, can show negative values and become unphysical. Moreover, the properties of the solutions are usually studied numerically for a given $b$, in a ``test-and-trial" approach that can be quite time expensive. Therefore, it is useful to have a general framework able to predict the qualitative features of the kinematical fields that will be produced by given choices of $b$ before solving the JEs; in this way one can constrain (as a function of the structural properties of the model under investigation) the values of $b$ that assure the positivity of $\vphiqm$, and produce the desired features of the azimuthal and radial velocity dispersion fields.

In a previous work (\citealt{2024MNRAS.530.1796D}, hereafter Paper I), we obtained general analytical results to address the problems mentioned above, also in the more general case of $b$  depending on $z$. The results were illustrated for the \cite{1975PASJ...27..533M} and \cite{1980PASJ...32...41S} disk models in the self-gravitating case, that allow for an almost complete analytical treatment; it turned out that in order to have a positive $\vphiqm$, upper limits exist on the $b$ values  and they depend mainly on the flattening of the disk, much less on the disk density profile; also, interestingly, these limits increase for increasing flattening. For a disk embedded in a dominant dark matter (DM) halo it is found that a round or near to round halo does not affect significantly the results obtained for the one-component models: again, the maximum value of $b$ corresponding to a positive $\vphiqm$ depends on the flattening of the stellar distribution, and increases for increasing flattening. 

With the present study we extend the preliminary investigation of Paper I to realistic galaxy models described as oblate ellipsoids, representative of the population of ``fast rotating" early-type galaxies (hereafter ETGs, see e.g. \citealt{Cappellari2016ARA&A..54..597C}). In particular, for the stellar component we adopt ellipsoidal generalizations of the $\gamma$-models and the S\'ersic models (ideal tools to explore systematic effects of different density profiles), also considering the presence of a DM halo with a DM-to-stellar mass ratio in the inner regions of the models similar to what estimated for real galaxies. With these models we can investigate quantitatively a few natural questions: 1) Is there any behavior of the critical $b$ values specifically associated with oblate shape and realistic density profiles, and that was missed by the analysis based on disk galaxy models used in Paper I? 2) In Paper I we considered dominant DM halos: for more realistic DM halos, how are the kinematical fields affected by the $b$-anisotropy, especially in the galaxy regions where the gravitational field switches from being stellar dominated to DM dominated? 3) Can we obtain some robust and sufficiently general indications about the dependence of the critical values of $b$ on the flattening and the density profile of the stellar component of the galaxy? 

The paper is organized as follows. In Section \ref{sec:jeans} we summarize the general relations established in Paper I to constrain the range of admissible values of $b$ for a given model. In Section \ref{sec:one_comp} we study one-component, self-gravitating oblate galaxy models, with density profiles described by the deprojected ellipsoidal S\'ersic law and by ellipsoidal $\gamma$-models. We explore how the limits on $b$ depend on galaxy flattening and on the specific density profile; illustrative examples of the kinematical fields for $b$ values near the critical limits are presented and discussed. In Section \ref{sec:two_comp} we consider a de Vaucouleurs oblate stellar density distribution coupled to a DM spherical halo, and we study how the critical $b$ values depend on the combined effects of the stellar density flattening and of the DM-to-stars mass ratio. Again, examples of the kinematical fields for $b$ values near the critical limits are presented. The main results are summarized in Section \ref{sec:disc_conc}. In the Appendix, two-component power-law ellipsoidal models are discussed in a fully analytical way, providing important hints to explain the results obtained numerically for the more realistic models discussed in the paper.


\section{The Jeans equations and the $B$, $C$, $D$ fields}
\label{sec:jeans}

In this Section we summarize the main results of Paper I, that are relevant for the present study. We consider axisymmetric galaxy models obeying the JEs in cylindrical coordinates
\begin{equation}
\begin{dcases}
    {\partial\rhos\sigz^2\over\partial z}=-\rhos {\partial\Phi\over\partial z}, \\\\
    {\partial\rhos\sigR^2\over\partial R}-{\rhos\Deltac\over R}=-\rhos
    {\partial\Phi\over\partial R}, \quad \Deltac\equiv\vphiqm -\sigR^2,
\end{dcases}
\label{eq:3int_jeans}
\end{equation}
where $\rhos(R,z)$ and $\Phi(R,z)$ are the stellar density distribution and the total (e.g., stars plus dark matter) gravitational potential, and $\sigz$ and $\sigR$ the vertical and radial velocity dispersions of the stellar component. A bar over a symbol indicates the average over velocity in phase-space, so that $\vphiqm =\vphim^2 +\sigphi^2$, where $\vphim$ is the streaming velocity field in the azimuthal direction, and $\sigphi$ is the azimuthal velocity dispersion. In particular, the only non-vanishing ordered velocity field is $\vphim$, and the off-diagonal terms of the velocity dispersion tensor vanish, i.e., the velocity dispersion ellipsoid is aligned at each point with the local coordinate basis, as in the classical two-integral case.

The solution of the first of equation (\ref{eq:3int_jeans}) is
\begin{equation}
    \rhos\sigz^2 = \int_z^\infty\rhos {\partial\Phi\over\partial z'}dz'.
\label{eq:2int_JE_sol}
\end{equation}
The radial JE equation is then solved for $\Deltac$ by assuming that $\sigR^2$ is linked to $\sigz^2$ through the choice of the function $b(z)\ge 0$, as
\begin{equation}
    \sigR^2 = b(z) \sigz^2,
    \label{eq:b_ans_def}
\end{equation}
so that for $b=1$ the solution reduces to the two-integral case. Following Paper I then
\begin{equation}
{\rhos\Deltac\over R} =b D + \rhos {\partial\Phi\over\partial R} =
b\Comm + (1-b)\rhos {\partial\Phi\over\partial R},
\label{eq:b_ans_delta}
\end{equation}
where
\begin{equation}
D\equiv {\partial\rhos\sigz^2\over\partial R}=\Comm -\rhos {\partial\Phi\over\partial R},
    \label{eq:d_def}
\end{equation}
and
\begin{equation}
\Comm\equiv \int_z^{\infty}\left(
{\partial\rhos\over\partial R}{\partial\Phi\over\partial z'}-{\partial\rhos\over\partial z'}{\partial\Phi\over\partial R}\right)\,dz'.
    \label{eq:jeans_rad_comm}
\end{equation}
Once $\Deltac$ is known
\begin{equation}
\vphiqm = \Deltac +b\sigz^2 = b B + R{\partial\Phi\over\partial R},
\label{eq:vphi2m_anis}
\end{equation}
where from the first identity in equation (\ref{eq:b_ans_delta})
\begin{equation}
B\equiv {R\over\rhos}D + \sigz^2.
\label{eq:b_def}
\end{equation}
Finally, from equation (\ref{eq:vphi2m_anis}):
\begin{equation}
\Delta\equiv\vphiqm -\sigz^2 =\Deltac +(b-1)\sigz^2 = b B + C,
\label{eq:delta_bans}
\end{equation}
where from equations (\ref{eq:d_def}) and (\ref{eq:b_def})
\begin{equation}
C\equiv R {\partial\Phi\over\partial R} -\sigz^2 = 
{R\over\rhos}\left[\rhos,\Phi\right] -B.
\label{eq:c_def}
\end{equation}
Notice that the functions $D$, $B$ and $C$ are independent of $b(z)$ and so for a given model they can be computed just once.

Models with $\Deltac\ge 0$ allow for a $k$-decomposition of $\vphiqm$ similar to that introduced in \cite{1980PASJ...32...41S}:
\begin{equation}
\vphim = k \sqrt{\Deltac}, \;\;\sigphi^2 = \sigR^2 +(1-k^2)\Deltac,
\label{eq:satoh_dec_bans}
\end{equation}
(\citealt{Cappellari2008MNRAS.390...71C}). 
Whenever $k^2<1$, one has $\sigphi^2 >\sigR^2$, i.e. the orbital anisotropy is tangential; radial anisotropy ($\sigphi^2 <\sigR^2$) requires $k^2>1$. We recall that in the \cite{1980PASJ...32...41S} decomposition and in its generalization to a spatially dependent $k$ (\citealt{Ciotti1996MNRAS.279..240C}) for two integral systems a formula identical to equation (\ref{eq:satoh_dec_bans}) holds, where $\Deltac$ is replaced by $\Delta$, and $\sigR$ by $\sigz$.

\subsection{Constraints on the $b(z)$ function}
\label{sec:phys_cons_bans}

Arbitrary choices of $b(z)$ can lead to unphysical solutions of the JEs, such as negative values of $\vphiqm$; information on the sign of $\Deltac$ and $\Delta$ is also relevant for the modeling because, although not directly related to model consistency, their positivity is needed to decompose $\vphiqm$ as in equation (\ref{eq:satoh_dec_bans}). As shown in Paper I and summarized below, it is possible to know, before solving the JEs, what constraints must be satisfied by $b(z)$ to avoid unphysical solutions, and also how specific choices of $b(z)$ affect the properties of the solutions in different regions of space. In the following, we restrict to the natural case of models with $\partial\Phi/\partial R\ge0$ everywhere.

\subsubsection{Positivity of $\vphiqm$}

The request of $\vphiqm\geq 0$ determines through equation (\ref{eq:vphi2m_anis}) two regions of the $(R,z)$-plane associated with the function $B$:
\begin{equation}
    \bpm = \left\{(R,z): B\gtrless 0\right\}.
    \label{eq:b_sets_def}
\end{equation}
We indicate with $\bzero$ the curve $B=0$ separating $\bminus$ and $\bplus$; with $\projz(\bpm)$ the {\it projection} of $\bpm$ on the $z$-axis, and finally with $\bzpm =\left\{R: B\gtrless 0, z\in\projz(\bpm)\right\}$ the {\it radial section} of $\bpm$ for given $z$ (see Figure 1 in Paper I for more details). $\vphiqm\ge 0$ over 
$\bplus$ independently of $b(z)$, while $\vphiqm \ge 0$ over $\bminus$ provided that at each $z\in\projz(\bminus)$,
\begin{equation}
b(z)\le\bM(z) \equiv
\min_{\bzminus} {R\over\lvert B\rvert}{\partial\Phi\over\partial R},
\label{eq:bmz_criterion}
\end{equation}
where the minimum is computed for $R$ spanning $\bzminus$.
Notice that when restricting the modeling to constant $b$, the condition above imposes $b\le\bM$, where $\bM$ is the minimum of $\bM(z)$.

\subsubsection{Positivity of $\Deltac$}

$\Deltac\geq 0$ is required to apply the $k$-decomposition to $\vphiqm$ in equation (\ref{eq:satoh_dec_bans}). The first identity in equation (\ref{eq:b_ans_delta}) determines the two sets
\begin{equation}
    \dpm =\left\{(R,z) : D\gtrless 0\right\},
    \label{eq:dpm}
\end{equation}
and $\dzero$ is the curve $D=0$ separating $\dplus$ and $\dminus$; $\Deltac\ge 0$ independently of $b(z)$ over
$\dplus$. The condition $\Deltac\ge 0$ over $\dminus$ requires instead that for $z\in\projz(\dminus)$
\begin{equation}
b(z)\le \bz(z) \equiv 
\min_{\dzminus}{\rhos\over\lvert D\rvert}{\partial\Phi\over\partial R}.
\label{eq:b0z_criterion}
\end{equation}
Again, in case of a constant $b$, the condition $\Deltac\ge 0$ reduces to $b\le \bz$, where $\bz$ is the minimum of $\bz(z)$. Notice that from equation (\ref{eq:b_def}) $\bminus\subseteq\dminus$, and $\dplus\subseteq\bplus$.

\subsubsection{Positivity of $\Delta$}
\label{sec:pos_delta}

The sign of $\Delta$ depends on the sets 
\begin{equation}
    \cpm = \left\{(R,z): C\gtrless 0\right\},
    \label{eq:cpm}
\end{equation}
separated by the curve $\czero$ where $C=0$. From Paper I,
\begin{enumerate}
    \item $\Delta\ge 0$ independently of $b(z)$ over $\bplus\cap\cplus$,
    over $\bplus\cap\cminus$ for
\begin{equation}
b(z)\ge\bu(z)\equiv\max_{\bzplus\cap\czminus}{\lvert C\rvert\over B},
\label{eq:b1_def}
\end{equation}
and over $\bminus\cap\cplus$ for
\begin{equation}
b(z)\le \bd(z)\equiv
\min_{\bzminus\cap\czplus}{C\over \lvert B\rvert}.
\label{eq:b2_def}
\end{equation}
    \item $\Delta < 0$ independently of $b(z)$ over $\bminus\cap\cminus$,
over $\bplus\cap\cminus$ for $b(z) < \bu(z)$, and over $\bminus\cap\cplus$ for $b(z)>\bd(z)$.
\end{enumerate}
In the case of a constant $b$, the critical values $\bu$ and $\bd$ are respectively the maximum of $\bu(z)$, and the minimum of $\bd(z)$.

\subsubsection{General rules}
\label{sec:gen_rules}

We recall here some general result that will be used in the following analysis. First, from Exercises 13.28-13.29 in C21, 
for an ellipsoidally stratified density $\rhos(m_*)$ in an ellipsoidally stratified potential $\Phi (m_{\rm h})$, the commutator $\Comm$ is everywhere positive (negative) when the surfaces $m_*$ are flatter (rounder) than the surfaces $m_{\rm h}$, and it vanishes if the density and potential flattenings are the same. $\Comm$ is also positive (negative) for self-gravitating oblate (prolate) ellipsoidal systems. 

Second, from equation (\ref{eq:2int_JE_sol}) it follows that a sufficient condition to have $D\leq 0$ over the whole $(R,z)$-plane is that $\partial\rhos/\partial R\leq 0$ and $\partial^2\Phi/\partial R\partial z\leq 0$ everywhere. Again from Exercise 13.29 in C21 this holds for any ellipsoidal $\rhos(m_*)$ in the potential produced by an ellipsoidal $\rhoh(m_{\rm h})$ (independently of the flattenings of $m_*$ and $m_{\rm h}$) if $d\rhos(m_*)/dm_*\leq 0$ and $d\rhoh(m_{\rm h})/dm_{\rm h}\leq 0$. Finally, from equation (\ref{eq:d_def}) it also follows that $D\leq 0$ everywhere for systems with $\Comm\leq 0$.

Third, from Paper I, if $\Comm <0$ everywhere then $\bplus\subseteq\cminus$, and $\cplus\subseteq\bminus$; if $\Comm\geq 0$ everywhere, then $\bminus\subseteq\cplus$, $\cminus\subseteq\bplus$, and 
\begin{equation}
\begin{dcases}
    \bu(z)\le 1,\quad\forall z\in\projz(\cminus),\\\\
    1\le \bz(z)\le \bd(z)\le \bM(z),\quad \forall z\in\projz(\bminus),
\end{dcases}
\label{eq:bz_hierarchy}
\end{equation}
whit the special case $\bminus =\cplus$, $\bplus =\cminus$, and $\bu=\bz=\bd=1$
if $\Comm =0$ everywhere.

We conclude by noticing that from these general results it follows that for all models considered in this paper, $\Comm\geq 0$ and $D\leq 0$ everywhere, with the consequences listed above.

\section{One-component models}
\label{sec:one_comp}

We investigate here how the limits on $b(z)$ depend on the stellar density profile, and on the galaxy flattening in one-component (i.e., no DM halo) 
ellipsoidal models; we also study how the choice of $b$ affects the kinematical 
fields resulting from the solution of the JEs. We consider ellipsoidal S\'ersic and $\gamma$-models, that are commonly used in the study of elliptical galaxies, as they reproduce well their stellar surface brightness profile. 
Thanks to the assumed ellipsoidal density stratification, the components $\partial\Phi/\partial R$ and $\partial\Phi/\partial z$ of the gravitational field needed for the solution of the JEs can be written as one dimensional integrals (e.g., see Exercise 2.6 in C21) that can be treated numerically in a very efficient way. In the following, the stellar mass-to-light ratio is assumed constant over the galaxy body. 

\subsection{Ellipsoidal S\'ersic models}
\label{sec:sérsic}

The edge-on projected surface density profile of an ellipsoidal \cite{1968adga.book.....S} model of index $n$, total stellar mass $\Ms$, and minor-to-major axial ratio $q$ is
\begin{equation}
    \Sigmas(R,z) = {\Ms\over\reff^2} {b_n^{2n}e^{-b_n\, l^{1/n}}\over \pi q\Gamma(2n+1)}, 
    \quad       l\equiv\sqrt{{R^2\over\reff^2} +{z^2\over q^2\reff^2}},
\label{eq:Sersic_brightness}
\end{equation}
where $\Gamma$ is the Euler complete gamma function,
\begin{equation}
    b_n = 2n - {1\over 3} + {4\over 405 n} + O\left(n^{-2}\right),
\end{equation}
(\citealt{1999A&A...352..447C}), and $\reff$ is the semimajor axis of the isophotal ellipse enclosing half of the total mass (luminosity) of the system. The deprojection $\rhos(m)$ of equation (\ref{eq:Sersic_brightness}), cannot be obtained in terms of elementary functions for generic $n$; here we adopt the very accurate analytical approximation in \cite{2025A&A...694A.118C}: for $n>1$
\begin{equation}
\rhos(m)={\Ms\over q\reff^3}{\cu\,{\rm e}^{-b_n\, m^{1/n}} m^{1/n-1}\over \left [1 + \cd^p\, m^{p/(2n)}\right]^{1/p}},\;\;m\equiv\sqrt{{R^2\over\reff^2} +{z^2\over q^2\reff^2}}
\label{eq:nua}
\end{equation}
where 
\begin{equation}
\cu = {b_n^{2n+1}\over 4\pi^2 n^2\Gamma (2n)}\, \mathrm{B}\left({1\over 2},{n-1\over 2n}\right),
\end{equation}
\begin{equation} 
\cd =\sqrt{{b_n\over 2\pi n}}\,\mathrm{B}\left({1\over 2},{n-1\over 2n}\right),
\label{eq:coeff}
\end{equation}
$\mathrm{B}$ is Euler complete beta function, and values of $p$ as a function of $n$ are provided in Table B.1 in that paper. In the following we consider the cases of $n=2,4,6$ (with $p(2)=1.718$, $p(4)=1.955$, $p(6)=2.021$), and $q=0.75,\,0.5,\,0.25$, for a total of 9 models; the $n=4$ case corresponds to the \cite{1948AnAp...11..247D} model (hereafter dV).

\begin{figure*}
\centering
        \includegraphics[width=0.9\linewidth]{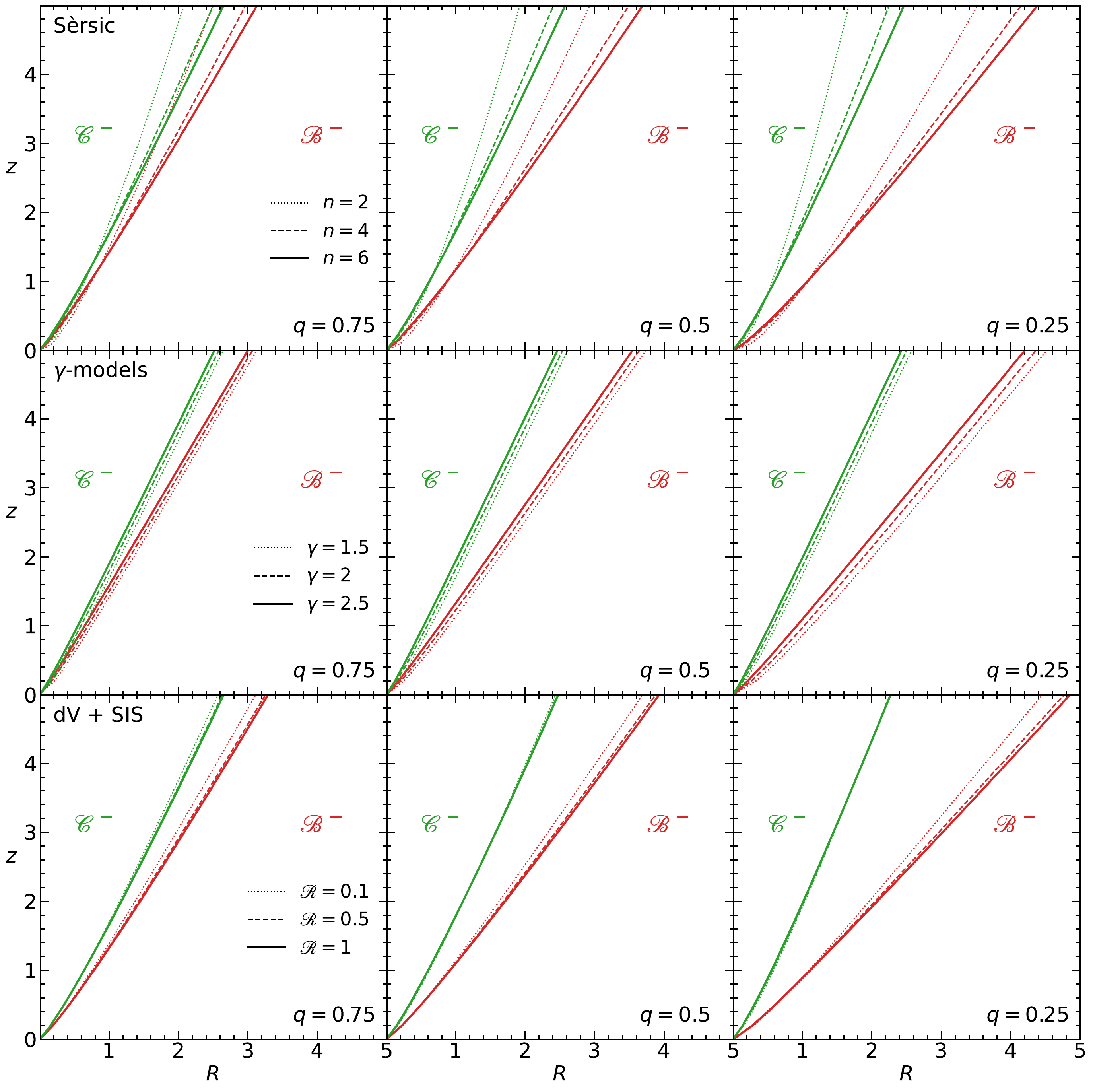}
        \caption{The ansatz-independent regions $\bpm$ and $\cpm$ for the ellipsoidal one and two-component models in Sections 3 and 4. For each model $\bminus$ is the part of the plane below $\bzero$ (red lines), and $\cminus$ is the part of the plane above $\czero$ (green lines); the stellar distribution flattening increases from left to right panels, with the axial ratio decreasing as $q=0.75$, $0.5$, and $0.25$. The different line styles identifies different values of the parameters of the models. Top panels: one-component S\'ersic model of index $n=2$, $4$, and $6$; $R$ and $z$ are in units of $\reff$, the semi-major axis of the effective isophote. Middle panels: one-component $\gamma$-models for $\gamma=1.5$, $2$, and $2.5$; $R$ and $z$ are in units of $\as$. Bottom panels: ellipsoidal stellar dV model of total mass $\Ms$in the potential of a SIS of circular velocity $\vh$, for different values of the parameter $\MR =\vh^2\reff/(G\Ms)$; $R$ and $z$ are in units of $\reff$.}
\label{f:fig1}
\end{figure*}

The $\bpm$ and $\cpm$ regions of the models are shown in Figure \ref{f:fig1} (top panels), with galaxy flattening increasing from the left to the right panel. The red lines (of increasing thickness for increasing $n$) show $\bzero$, the separation line between $\bminus$ (the region below $\bzero$, where $\vphiqm$ and $\Delta$ can become negative somewhere, respectively for $b>\bM$ and $b>\bd$), and $\bplus$ (the region above $\bzero$, where $\vphiqm$ increases for increasing $b$). It is apparent how $\bminus$ shrinks toward the equatorial plane for increasing $n$ at fixed $q$, and for increasing galaxy flattening at fixed $n$, with the effect of flattening dominant over that of the density profile. The green lines (of increasing thickness for increasing $n$) show $\czero$, the separation line between $\cminus$ (the part of the plane above $\czero$, where $\Delta$ can become negative somewhere for $b<\bu$), and $\cplus$ (the part below $\czero$): in agreement with the general rules for models with positive commutator, $\bminus\subseteq\cplus$ and $\cminus\subseteq\bplus$. The region $\cminus$ expands for increasing $n$ at fixed $q$, but it contracts toward the symmetry axis for increasing flattening at fixed $n$; this is particularly evident for the $n=2$ model. From the figures, it is also apparent how for decreasing flattening (i.e., rounder shapes), the red and green lines tend to become nearer, in agreement with the fact that in case of null commutator (as in spherical models), $\bzero=\czero$. Finally, it should be noticed how these lines are almost straight for all $n$ and $q$. All these properties can be understood by using the simple power-law models described in the Appendix: in particular, the left panels of Figure \ref{f:figA1} illustrate the dependence of the $\bzero$ and $\czero$ curves on the model flattening.

\begin{figure*}
\centering
        \includegraphics[width=0.7\linewidth]{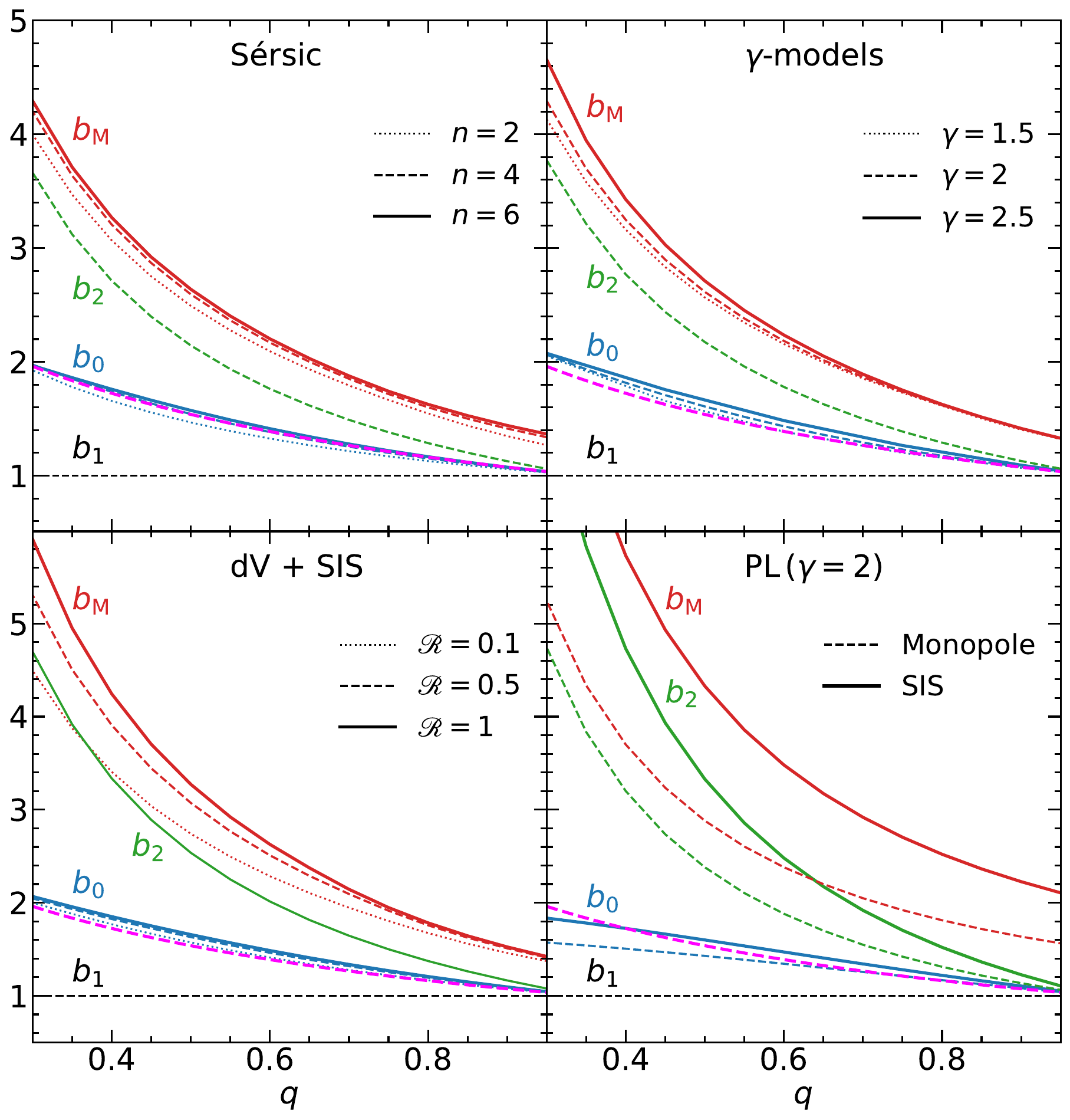}
        \caption{The critical values $\bu$ (black), $\bz$ (blue), $\bd$ (green), and $\bM$ (red), as a function of the axial ratio $q$ of the stellar component (rounder models for increasing $q$), for the same one-component S\'ersic models (top left), $\gamma$-models (top right), and dV models in a SIS DM halo (bottom left) shown in Figure \ref{f:fig1}. Bottom right: the critical $b$ curves for the $\gamma=2$ power-law ellipsoidal stellar model in a dominant SIS potential (thick continuous lines, Appendix \ref{sec:PL_SIS}), and in a dominant monopole potential (thin dashed lines, Appendix \ref{sec:pl_mon}). For simplicity, the $\bd$ green lines are shown only for the dV, the $\gamma=2$, and the dV+SIS ($\MR = 0.5$) models. As discussed in Section 5, the magenta line represents the limit on the $b$-anisotropy as a function of galaxy flattening determined by \protect\cite{Cappellari2007MNRAS.379..418C}.}
\label{f:fig2}
\end{figure*}

Having identified the $\bpm$ and $\cpm$ regions, we can now discuss the critical values of $b$ in terms of $n$ and $q$. These are shown in the top left panel of Figure \ref{f:fig2}, with red, green, blue and black dotted lines corresponding respectively to $\bM$, $\bd$, $\bz$, and $\bu$. We recall that $\bM$ is related to the sign of $\vphiqm$, with $b >\bM$ leading to an unphysical $\vphiqm <0$ somewhere in $\bminus$, while $b<\bz$ guarantees that $\Deltac >0$ everywhere; finally, $\Delta >0$ everywhere for $\bu<b<\bd$. The first general result is that in rounder models all critical values of $b$ decrease, i.e., a lower $b$-anisotropy can be imposed, while flatter density distributions can support larger anisotropies. The second important result is that the density profile has a minor effect on the critical $b$ values, similar to what found for the position and extension of the $\bpm$ and $\cpm$ regions; in fact, for simplicity and for its special relevance, the $\bd$ line was shown only for the $n=4$ (dV) model. Again, these results can be interpreted with the the models in Appendix, whose critical $b$ can be expressed analytically, and that are shown in the bottom right panel of Figure \ref{f:fig2}; in particular, the comparison should be carried out with the thin lines, corresponding to models in a monopole potential, the situation holding naturally at intermediate/large radii in finite mass models. It is interesting to see how not only qualitatively, but also quantitatively the power-law toy-model lines reproduce those of the S\'ersic models, even though these last ones are not power-laws.

\begin{figure*}
\centering
        \includegraphics[width=0.8\linewidth]{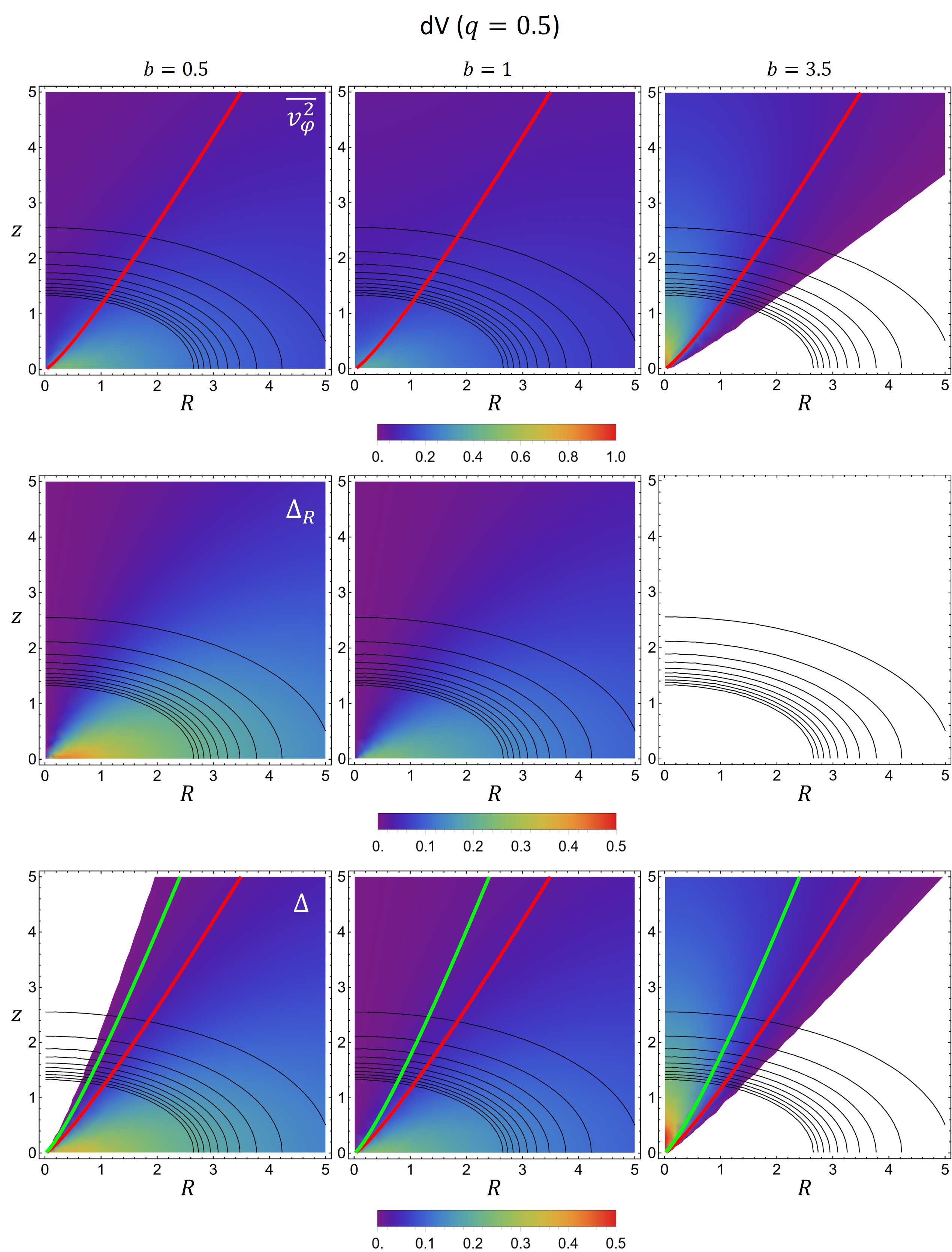}
        \caption{From top to bottom, maps in the $(R,z)$-plane of $\vphiqm$, $\Deltac$, and $\Delta$ for the dV model with $q=0.5$; all the fields are in units of $G\Ms/\reff$, $R$ and $z$ in units of $\reff$. Columns correspond (from left to right) to $b=0.5$, $1$, and $3.5$. Red and green curves represent the $\bzero$ and $\czero$ lines, as in Figure \ref{f:fig1}; black ellipses are equally spaced stellar isodensity contours. In the white regions, the value of the field is negative.}
\label{f:fig3}
\end{figure*}

\begin{figure*}
\centering
        \includegraphics[width=0.8\linewidth]{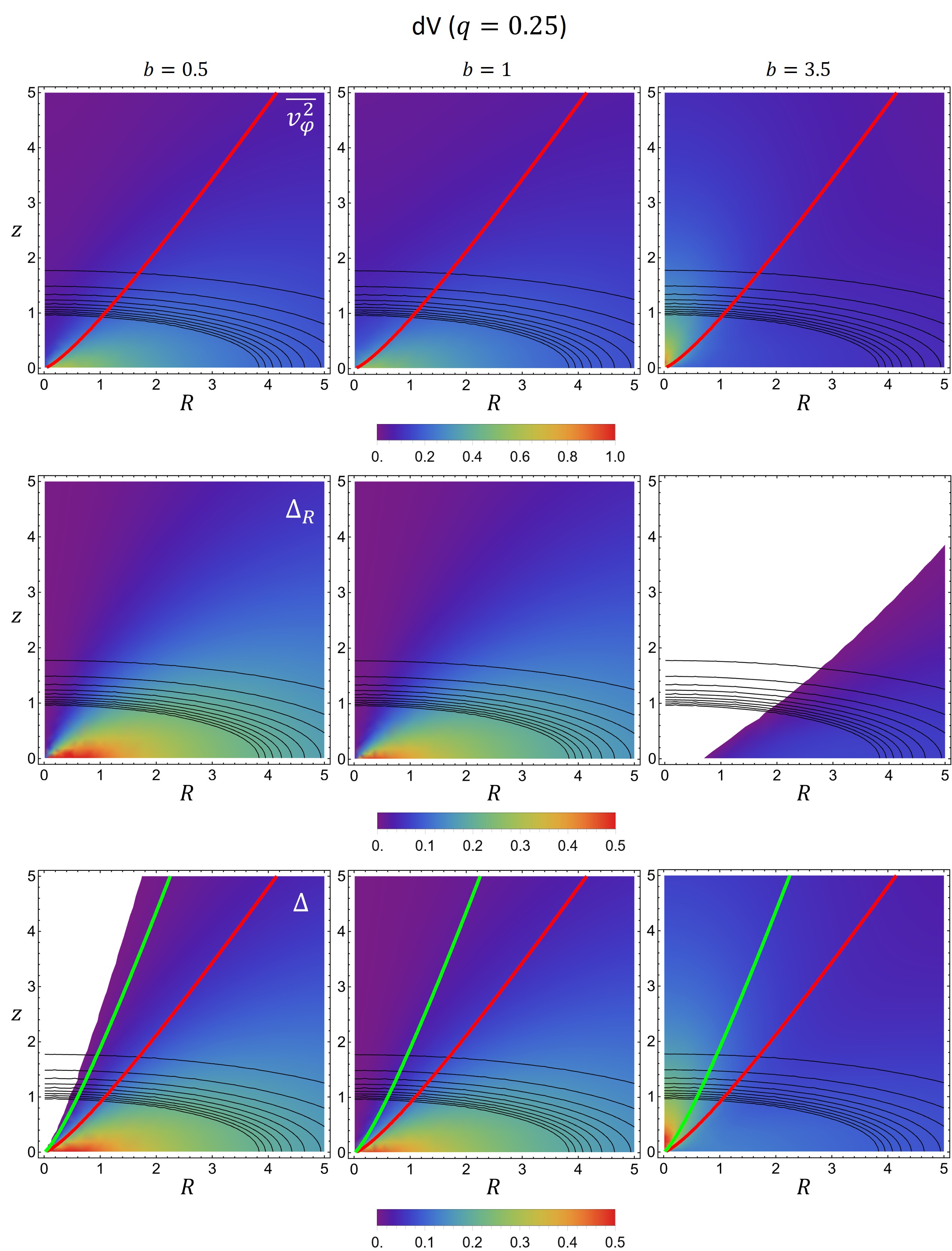}
        \caption{Same as in Figure \ref{f:fig3}, but with an increased flattening of the stellar distribution, corresponding to the axial ratio of $q=0.25$.}
\label{f:fig4}
\end{figure*}

Figures \ref{f:fig3} and \ref{f:fig4} illustrate the effects of different choices of $q$ ($0.5$ and $0.25$, respectively), and $b$ ($0.5$, $1$, and $3.5$, from the left to the right columns) on the fields $\vphiqm$, $\Deltac$ and $\Delta$ of the one-component dV model; the $b=1$ case corresponds to a classical two-integral phase-space distribution function. Following Figure \ref{f:fig1}, the red and green lines separate the $\bpm$ and $\cpm$ regions. For the less flattened model (Figure \ref{f:fig3}), as $b$ increases, $\vphiqm$ (top panels) increases over $\bplus$ and decreases in $\bminus$, and it is negative when $b=3.5$, being this value higher than the critical value $\bM$ (see the top left panel in Figure \ref{f:fig2} for $q=0.5$). Analogously, $\Deltac$ (middle row) decreases over all the plane for increasing $b$, and for $b=3.5$ it is negative everywhere, again in accordance with the value of $\bz$ in Figure \ref{f:fig2}. The behavior of the $\Delta$ field in the bottom row is more complicated, as it depends on the value of $b$ with respect to $\bu$ and $\bd$: as expected, $\Delta$ is then found negative over some part of $\cminus$ when $b<\bu=1$ (left panel), and negative over some part of $\bminus$ for $b>\bd$ (right panel), while in the two integral case (middle panel) we have $\Delta=\Deltac=\Comm\geq 0$ everywhere.

The effect of increasing the density flattening from $q=0.5$ to $q=0.25$ is illustrated by Figure \ref{f:fig4}. The effects are quite significant, and in line with the expectations. In fact, from Figure \ref{f:fig2}, the values of the critical $b$ increase for increasing flattening; thus the same $b=3.5$ adopted in Figure \ref{f:fig3} corresponds now to $\vphiqm$ and $\Delta$ everywhere positive, and to the negative region for $\Deltac$ less extended than in Figure \ref{f:fig3}.


\subsection{Ellipsoidal $\gamma$-models}

We now consider the family of one-component ellipsoidal $\gamma$-models (\citealt{1993MNRAS.265..250D}, \citealt{1994AJ....107..634T}) of total mass $\Ms$, axial ratio $q$, scale length $\as$, and inner density slope $0\leq\gamma <3$:
\begin{equation}
    \rhos(m) ={\Ms\over\as^3}{3-\gamma\over 4\pi q m^{\gamma}(1 + m)^{4-\gamma}},\;
    m =\sqrt{{R^2\over\as^2} + {z^2\over q^2\as^2}}.
\end{equation}
In the central region $\rhos\sim m^{-\gamma}$, therefore these models are somewhat complementary to S\'ersic models: they allow to study the effects of significant changes of the density slope in the central region, where for S\'ersic models with $n>1$ the density is $\propto m^{1/n-1}$, and so it cannot be very steep even for very large values of $n$ (see equation \ref{eq:nua}, see also \citealt{1991A&A...249...99C}). Instead, in the external regions (where the density slope of S\'ersic models changes as a function of the distance from the center) the $\gamma$-models are an almost perfect power-law $\rhos\sim m^{-4}$ independently of $\gamma$, and so we expect that their behavior is captured with high accuracy by the pure power-law models in the Appendix.

In the middle row of Figure \ref{f:fig1} we show the critical regions for different values of $q$ (with flattening increasing from the left panel to the right panel), and for $\gamma = 1.5$, $2$, and $2.5$ (with line thickness increasing with $\gamma$). The similarity of the $\bpm$ and $\cpm$ regions with those of S\'ersic models of same $q$ is quite remarkable, and again $\bminus$ and $\cminus$ shrinks (with the $\bzero$ and $\czero$ lines rotating respectively towards the equator, and the symmetry axis) for increasing flattening; notice how the corresponding red and green lines (outside the core radius $\as$) are almost perfectly straight, in agreement with the results in Appendix \ref{app:pl_models}, where it is shown that in fact this is what happens for pure power-law ellipsoidal densities. Finally, also for these models the effect of the density profile is much less important than that of the flattening.

The critical values of $\bM$ and $\bz$ for $\gamma =1.5$, $2$, and $2.5$, as a function of the axial ratio $q$, are shown in the top right panel of Figure \ref{f:fig2}, together the $\bd$ line for the $\gamma=2$ model; in remarkable similarity with the case of the ellipsoidal S\'ersic models, rounder shapes can allow for lower $b$-anisotropy, while the effects of the specific density profile are much less important. Due to the power-law nature of the density of $\gamma$-models outside the core region, the results are reproduced very well by the models in Appendix \ref{app:pl_models}, especially by the model with monopole potential in Appendix \ref{sec:pl_mon}, as apparent from the bottom right panel in Figure \ref{f:fig2} (thin lines).

The fields $\vphiqm$, $\Deltac$ and $\Delta$ of $\gamma$-models, for the same values of $q$ and $b$ used for S\'ersic models, are so similar to the fields in Figures \ref{f:fig3} and \ref{f:fig4}, including for the white negative regions, that we do not show them; we stress however that this fact reinforces the conclusion that flattening seems much more important than the specific density profile in determining the effects of the $b$-anisotropy on the kinematical properties, at least for the very general class of models analyzed in this work.

\section{An ellipsoidal de Vaucouleurs galaxy in a Singular Isothermal Sphere DM halo}
\label{sec:two_comp}

Here we finally explore how the presence of a DM halo affects the anisotropy constraints 
and what are the effects of $b$ on the kinematical fields, similarly to what discussed in Section \ref{sec:one_comp}. In particular, we study an ellipsoidal \cite{1948AnAp...11..247D} stellar density of total mass $\Ms$, effective radius $\reff$, and axial ratio $q$, described by equation (\ref{eq:Sersic_brightness}) with $n=4$, embedded in a DM halo described by a Singular Isothermal Sphere (SIS) of circular velocity $\vh$ and potential
\begin{equation}
    \phih =\vh^2\ln {r\over\reff},\quad r=\sqrt{R^2+z^2}.
\label{eq:phi_sis}
\end{equation}
At variance with the two-component models in Appendix (and in Paper I), where the stellar component of the galaxy is assumed to be a tracer in a dominant DM potential, in the present dV+SIS model the total potential is
\begin{equation}
    \Phi = \Phi_* + \MR\ln r,\quad \MR\equiv {\vh^2\reff\over G\Ms},
\end{equation}
where $\Phi_*$ is the potential of the stellar component. In the expression above $\Phi$ is normalized to $G\Ms/\reff$, and the lengths to $\reff$; the dimensionless coefficient $\MR$ therefore measures the importance of the DM potential relative to the stellar one: at variance with the halo-dominated models, we can now explore also the transition region between the stellar dominated inner region and the outer, DM halo dominated region. 

The range of values for $\MR$ is chosen from the (projected) DM-to-stellar mass ratio inside $\reff$ for a spherical dV\footnote{Of course, equation (\ref{eq:MRproj}) applies also to a generic spherical S\'ersic profile of index $n$, coupled to a SIS, for which
\begin{equation}
\Sigma_{\rm DM}(R)={\vh^2\over 4 G R},\quad M_{\rm p,DM}(R)={\pi \vh^2 R\over 2 G}.
\end{equation}
} stellar component:
\begin{equation}
{M_{\rm p,DM}(\reff)\over M_{\rm p,*}(\reff)}=\pi\MR.
\label{eq:MRproj}
\end{equation}
We consider the range $0\leq \MR\leq 1$, even though realistic values for $\MR$ would be around $0.1-0.2$ (e.g., \citealt{2017MNRAS.467.1397P}); note that the DM-to-stellar mass ratio within spheres in the three dimensional space instead of circles in the projection plane, would give values for $\MR$ very similar to those in the equation above.

\begin{figure}
\centering
        \includegraphics[width=0.95\linewidth]{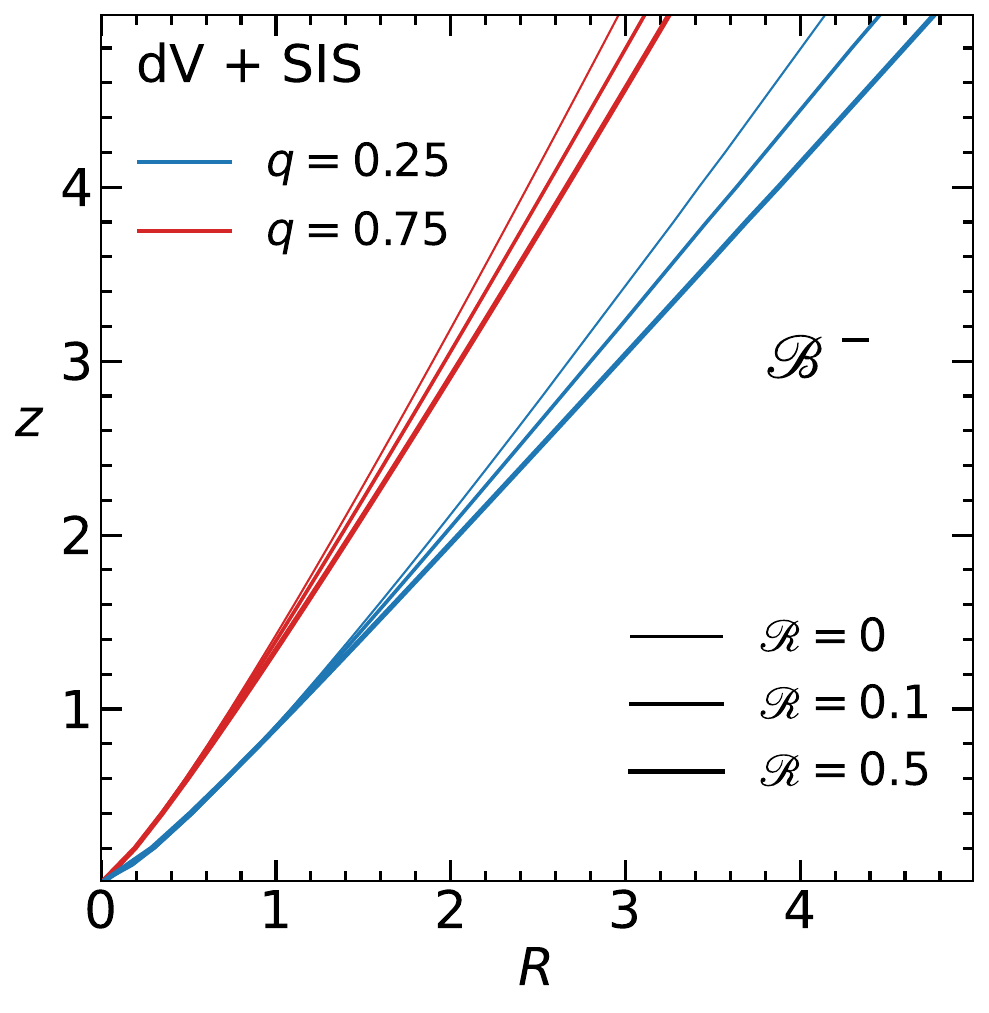}
        \caption{The $\bpm$ regions of the one and two-component dV models for two flattenings of the stellar distribution (red and blue lines), and three values of the DM-to-stellar mass ratios for each flattening (lines of increasing thickness). As in Figure \ref{f:fig1}, where these curves are present, $\bminus$ for each models lies below the corresponding line. $R$ and $z$ are in units of $\reff$. The figure illustrates how stellar flattening affects the size of the regions more than the DM presence, even for dominant halos.}
\label{f:figNEW}
\end{figure}

In the three bottom panels of Figure \ref{f:fig1} we show the $\bzero$ (red) and $\czero$ (green) lines for increasing flattening of the stellar density distribution, and for three values of $\MR$; for comparison, the no-halo case is given by the $n=4$ curves in the three top panels. Qualitatively, the behavior of the dV+SIS models is very similar to that of the one-component dV models of same flattening. Again, the most significant parameter determining the position and size of the $\cpm$ and $\bpm$ regions appears to be the flattening of the stellar density distribution, rather than the DM-to-stellar mass ratio. In order to better illustrate this result, Figure \ref{f:figNEW} shows in the same panel a few curves of Figure \ref{f:fig1} delimiting the $\bpm$ regions for dV models two different flattenings, and three different $\MR$ values. The only effect of a DM halo (even a very massive one) is to moderately reduce the size of the $\bminus$ region. The DM effect is however larger than that due to a change in the stellar density profile in one-component models.

In Figure \ref{f:fig2} (bottom left panel) we show the critical values of $b$ as a function of $q$, for the three values of $\MR$ previously considered; again, the $\MR=0$ case to be used for comparison is shown by the $n=4$ curves in the top left panel. As for the one-component models, more and more spherical stellar densities are less and less able to sustain significant $b$ anisotropy; moreover, at fixed flattening an increasing amount of DM appears to (slightly) increase the range of values admissible for $b$. These two findings also hold for the DM halo-dominated power-law stellar models of Appendix \ref{sec:PL_SIS} (heavy solid $b$ curves in the bottom right panel to be compared with the $\gamma=2$ curves in the top right panel): the effect of a DM halo is more important than that of the specific stellar density profile, but lower than that of the flattening.

\begin{figure*}
\centering
        \includegraphics[width=0.8\linewidth]{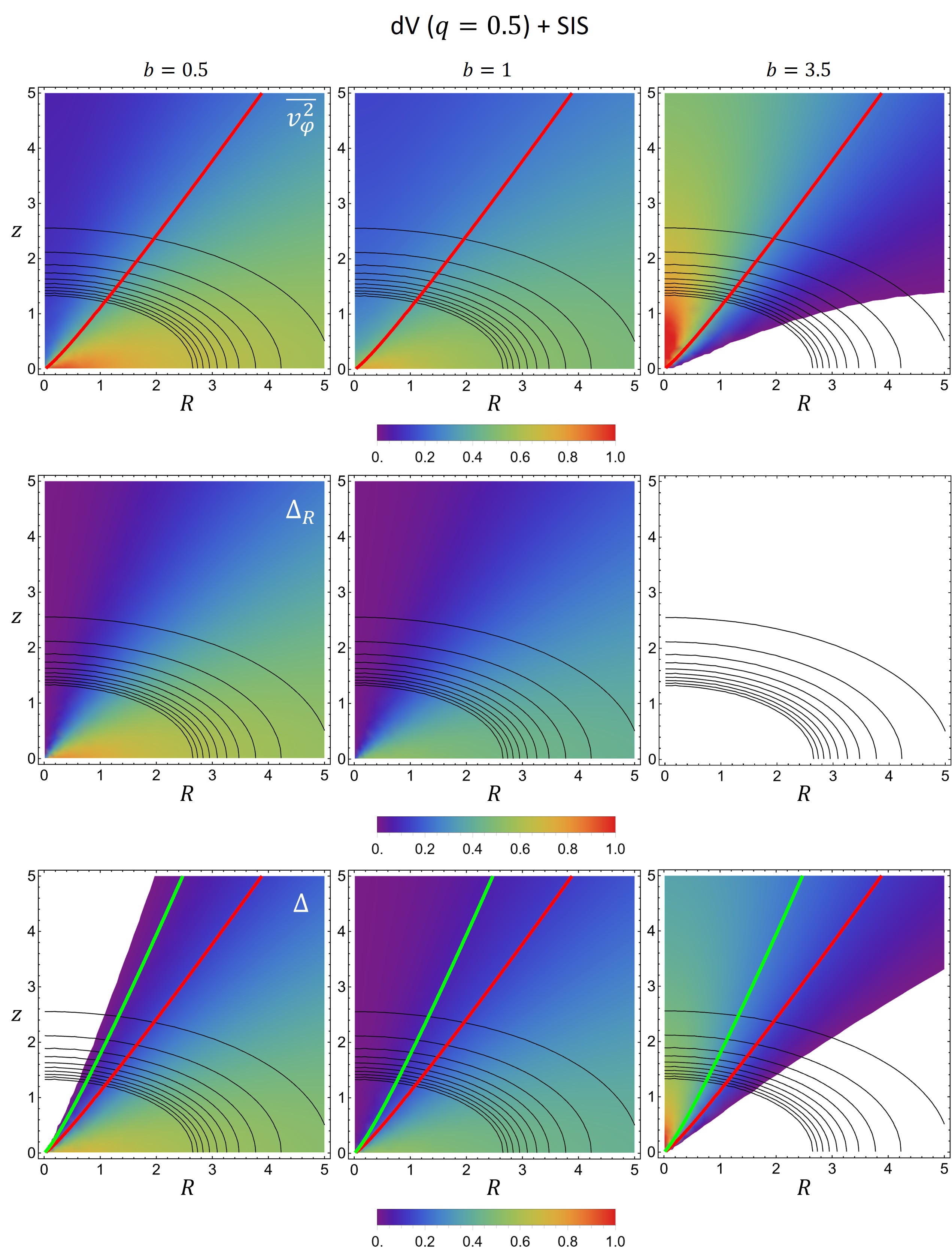}
        \caption{Maps in the $(R,z)$-plane of $\vphiqm$ (top row), $\Deltac$ (middle row) and $\Delta$ (bottom row) in units of $G\Ms/\reff$, for an ellipsoidal dV+SIS model with $q=0.5$ for the stellar component, and $\MR =0.5$ for the SIS DM halo, for $b=0.5$ (left column), $b=1$ (central column) and $b=3.5$ (right column). Black solid lines represent equally spaced stellar isodensity contours. The red and green lines show, respectively, the $\bzero$ and $\czero$ curves, and white regions correspond to negative values of the fields.}
\label{f:fig5}
\end{figure*}

\begin{figure*}
\centering
        \includegraphics[width=0.8\linewidth]{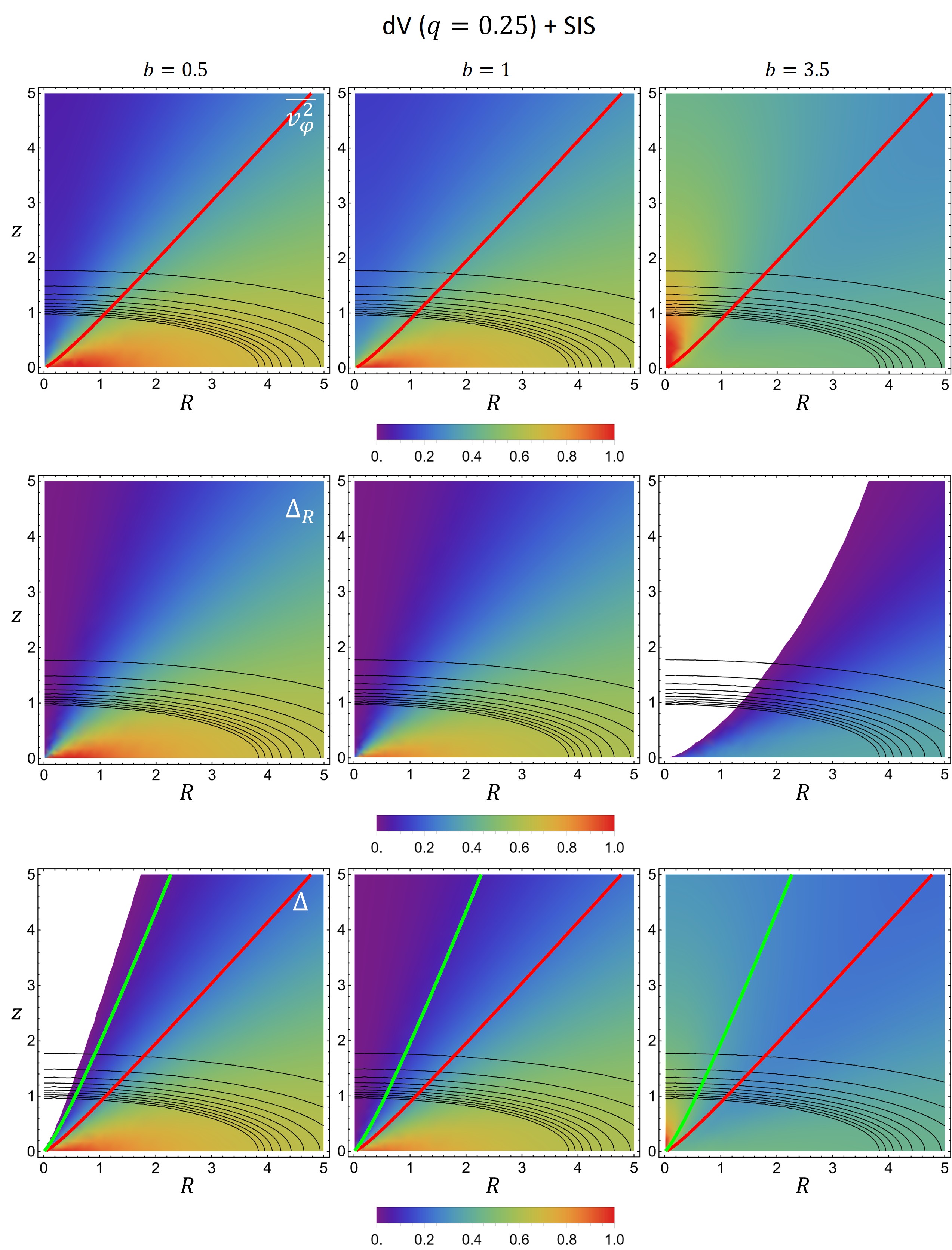}
        \caption{Same as in Figure \ref{f:fig5}, but with an increased flattening of the stellar distribution, corresponding to the axial ratio of $q=0.25$.}
\label{f:fig6}
\end{figure*}

Finally, in Figures \ref{f:fig5} and \ref{f:fig6} we show the fields $\vphiqm$, $\Deltac$ and $\Delta$ for the dV+SIS models with $\MR=0.5$, for increasing stellar flattening ($q=0.5$ and $q=0.25$) and for three constant values of $b$. The effect of the DM halo is apparent by comparison with Figures \ref{f:fig3} and \ref{f:fig4}, that show the one-component dV models, with same flattening and $b$ values. Of course, as shown by the color scales, the absolute values of the fields are substantially increased by the presence of DM. Consistently with the increase of the critical $b$ values produced by the DM halo, for $b=3.5$ and $q=0.5$, the region where $\vphiqm$ of the dV+SIS model is negative is smaller than that of the one-component model; and similarly, for $q=0.25$ and $b=3.5$, the region where $\Deltac <0$ becomes smaller for the dV+SIS then for the dV models.


\section{Discussion and conclusions}
\label{sec:disc_conc}

In a follow-up of Paper I (\citealt{2024MNRAS.530.1796D}), in the present work we studied how the general results obtained there on the limits on the $b$-anisotropy, and on its effects on the kinematical fields of axisymmetric systems, apply to one and two-component ellipsoidal galaxy models with realistic density profiles. In fact, for analytical simplicity and illustrative purposes, the models adopted in Paper I were the \cite{1975PASJ...27..533M} and \cite{1980PASJ...32...41S} disks, that especially in the case of low flattening are quite different from real ETGs.

In particular, the present study investigates how the constraints on $b$ (for example those imposed by the positivity of $\vphiqm$ or of $\Deltac =\vphiqm -\sigR^2$) depend on the stellar density profile, flattening, and DM amount and distribution; we consider oblate models with a stellar component of axial ratio $0<q<1$, avoiding the more complicated prolate systems. For the stellar profiles, we consider the ellipsoidal generalization of the widely used S\'ersic and $\gamma$-models. The two-component galaxies are made by de Vaucouleurs stellar ellipsoids (i.e., $n=4$ S\'ersic models) of different $q$ coupled with Singular Isothermal Sphere DM halos of adjustable circular velocity. The Appendix presents three simple models consisting of a power-law stellar ellipsoid embedded in three different DM halo-dominated potentials (a logarithmic ellipsoidal potential of same flattening as the stellar distribution, a SIS potential, and a monopole potential). These models albeit very idealized allow for a complete analytical study, and have been used to explain the numerical results obtained for the more realistic models. 

The main results are summarized as follows:
\begin{itemize}

    \item For all the models considered in this work $\partial\Phi/\partial R\geq 0$, so that $\vphiqm\geq 0$ over $\bplus$ independently of the value of $b$, while $\vphiqm$ becomes negative over larger parts of $\bminus$ for increasing $b>\bM$. Moreover, $\Comm\geq 0$ and $D\equiv \partial\rhos\sigz^2/\partial R\leq 0$ over the whole $(R,z)$-plane.
 
    \item From the previous properties, the following results descend: 1) from the positivity of the commutator, $\bminus$ is contained in $\cplus$, $\cminus$ in $\bplus$, and the critical values of $b$ are ordered as $\bu\leq 1\leq\bz\leq\bd\leq\bM$; 2) from the negativity of $D$, $\Deltac=\vphiqm-\sigR^2$ becomes negative over increasing portions of the $(R,z)$-plane for increasing $b>\bz$; 3) $\Delta =\vphiqm-\sigz^2$ is positive over $\bplus\cap\cplus$ independently of $b$, over $\cminus$ for $b>\bu$, and over $\bminus$ for $b<\bd$.

    \item The regions of the $(R,z)$-plane where $\vphiqm$, $\Deltac$, and $\Delta$ become negative, for $b$ values failing to satisfy the constraints, have remarkably similar shapes for the different ellipsoidal (one and two-component) models here explored; these shapes are also similar to those found for the disk models in Paper I. In particular $\bminus$ contains the equatorial plane, and $\cminus$ includes the $z$-axis. 
    
    \item Both for one and two-component models, the flattening of the stellar density has the most relevant effect in determining the critical values of $b$, and the size and shape of the $\bpm$ and $\cpm$ regions. Flatter models are able to sustain larger $b$-anisotropy (i.e., all the critical $b$ values increase for increasing galaxy flattening), and their $\bm$ regions shrinks toward the equatorial plane. In two-component models, a more massive DM halo also allows for a larger anisotropy, but the effect is smaller than that due to the galaxy flattening. Finally, the specific stellar density profile appears to have a very minor effect in determining the critical $b$ values and the shape the critical regions.
    
\end{itemize}

We conclude with a comment on the relation of our work with a decomposition for the azimuthal motions alternative to that of \cite{1980PASJ...32...41S}, and based on the parameter $\gammap$ defined as 
\begin{equation}
    \gammap\equiv 1-{\sigphi^2\over\sigR^2}\leq 1,
    \label{eq:gamma_def}
\end{equation}
so that $\gammap>0$ corresponds to radial anisotropy, and $\gammap <0$ to tangential anisotropy (\citealt{Cappellari2007MNRAS.379..418C}); in two-integral systems, $\gammap =0$ corresponds to the isotropic case. At each point of the system $\vphim^2 =\Deltac +\gammap\sigR^2$, and from the positivity of $\vphim^2$ one has that
\begin{equation}
-{\Deltac\over\sigR^2}\le\gammap\le 1.
\label{eq:gamma_cons}
\end{equation}
In case of spatially constant $\gammap$, its minimum value over the $(R,z)$ plane is $\max (-\Deltac/\sigR^2)$. In general $\gammap$ can be either positive or negative where $\Deltac >0$, but it can only be positive where $\Deltac <0$; in particular, in order to have globally ``$\varphi R$-isotropic'' ($\sigphi=\sigR$ and $\gammap=0$)  galaxies, $\Deltac$ must be positive everywhere, i.e., $b\leq\bz$. As for increasing flattening (and to a lesser extent for increasing DM-to-stellar mass ratio) $\bz$ increases, it follows that 
more flattened stellar densities and/or higher DM-to-stellar mass ratios can be ``$\varphi R$-isotropic'' and at the same time more ``$z R$-anisotropic" than less flattened stellar densities, and/or lower DM-to-stellar mass ratios.

Remarkably, by using the JAM method, it has been shown that the stellar kinematics of regular rotators in the ATLAS$^{\rm 3D}$ survey can be successfully modeled on average with $b>1$ and $\gammap\simeq 0$ within about one effective radius (e.g., \citealt{Cappellari2016ARA&A..54..597C}). The orbital anisotropy turned out to be limited by $1-1/b<0.7\varepsilon$, where $\varepsilon$ is the intrinsic galaxy flattening, i.e. $b<1/(0.3+0.7q)$. This limit had first been found with the Schwarzschild orbit superposition method for the galaxies in the SAURON survey (\citealt{Cappellari2007MNRAS.379..418C}), and thereafter has been often confirmed for larger samples of regular rotators, as those of the MaNGA survey (e.g., \citealt{Wang2020MNRAS.495.1958W}). We notice that the condition $b<\bz(q)$ discussed above to have $\gammap=0$, coupled with 
the fact that $\bz(q)$ is almost independent of the specific stellar density profile, provide a natural explanation for the existence of the above empirical limit on $b$. Indeed, there is a remarkable similarity of the magenta line (the empirically determined limit on $b$ as a function of $q$) in Figure \ref{f:fig2} with the $\bz(q)$ lines. 
It would be interesting to extend the comparison of the present results to the galaxies obtained by high-resolution numerical simulations of galaxy formation in a cosmological setting, such as the IllustrisTNG suite (see e.g. \citealt{2018MNRAS.473.4077P}, \citealt{2019MNRAS.490.3196P}, \citealt{2019ComAC...6....2N}). The spatial resolution of such simulations is large enough to check whether the $b$-ansatz can be applied (for example, whether the velocity dispersion tensor is aligned with the local basis of coordinates), how well the $b$-anisotropy (in case for a spatially dependent $b$) can describe the kinematical fields, and whether the $b$ values are related with the formation history of the simulated galaxies, their environment, and the relative importance of the DM halo.

\section*{Acknowledgments}
We thank the Referee, Eduardo Vitral, for a careful reading of the manuscript, and for insightful comments. LC and SP acknowledge support from the Project PRIN MUR 2022 (code 2022ARWP9C) ‘Early Formation and Evolution of Bulge and HalO (EFEBHO)’, PI: M. Marconi, funded by European Union – Next Generation EU (CUP J53D23001590006). LDD acknowledges the support from the International PhD College - Collegio Superiore, University of Bologna, 40129 Bologna, Italy.

\section*{Data availability}

No datasets were generated or analyzed in support of this research.

\bibliographystyle{mnras}
\bibliography{dd25}

\appendix
\section{Power-law ellipsoidal galaxy models}
\label{app:pl_models}

We present here three very idealized two-component models that allow for a fully analytical treatment, and illustrate in a simple way the results obtained numerically for the more realistic models in Sections \ref{sec:one_comp} and \ref{sec:two_comp}. Also for the models The considerations in Section \ref{sec:gen_rules} show that also for these models $\Comm\geq 0$ and $D\leq 0$ everywhere. Their stellar component is an ellipsoidal, power-law density profile of axial ratio $q$ and density slope $0<\gamma<3$
\begin{equation}
    \rhos = {\rho_0\over q m^{\gamma}},\quad m \equiv\sqrt{R^2 + {z^2\over q^2}};
    \label{eq:pl_ell_rho}
\end{equation}
in all the Appendix, lengths are intended normalized to some scale length $\as$, and densities to $\rho_0$.

\subsection{Power-law ellipsoidal galaxies in a dominant logarithmic potential of same flattening}
\label{sec:PL_Log}

In this first family the stellar density in equation (\ref{eq:pl_ell_rho}) is embedded in a dominant DM halo with ellipsoidal logarithmic potential of same flattening, and circular velocity $\vh$:
\begin{equation}
    \phih =\vh^2\ln m;
\label{eq:phi_sis}
\end{equation}
in the following, all the velocities are in units of $\vh$.
From equations (\ref{eq:2int_JE_sol}) and (\ref{eq:jeans_rad_comm}),
\begin{equation}
\rhos\sigz^2 ={q^{\gamma-1}\over \gamma (q^2R^2+z^2)^{\gamma/2}},\qquad
\left[\rhos,\phih\right] = 0,
\end{equation}
with $\sigz^2 = 1/\gamma$, independently of $q$. From equations (\ref{eq:d_def}),  (\ref{eq:b_def}), and (\ref{eq:c_def})
\begin{equation}\label{eq:pllog_D}
D=-{q^{\gamma+1}\over R^{1+\gamma}(q^2+s^2)^{\gamma/2 +1}},\qquad s\equiv{z\over R},
\end{equation}
\begin{equation}\label{eq:pllog_BC}
B={(1-\gamma) q^2 +s^2\over\gamma (q^2 + s^2)}= -C,
\end{equation}
so that the $B$ and $C$ functions are constant on straight lines $z =s R$ of constant $s$, and from the vanishing of the commutator $\bminus =\cplus$, $\bplus = \cminus$. For $\gamma < 1$ the regions $\bplus$ and $\cminus$ coincides with the whole $(R,z)$-plane, while for $\gamma\geq 1$, $\bminus$ is the portion of the $(R,z)$-plane below the line
\begin{equation}
z = q\,\sqrt{\gamma - 1} R,
\label{eq:pl_log_bminus}
\end{equation}
so that $\bminus$ expands both for increasing $q$ and increasing $\gamma$. 

\begin{figure*}
\centering
\includegraphics[width=0.7\linewidth]{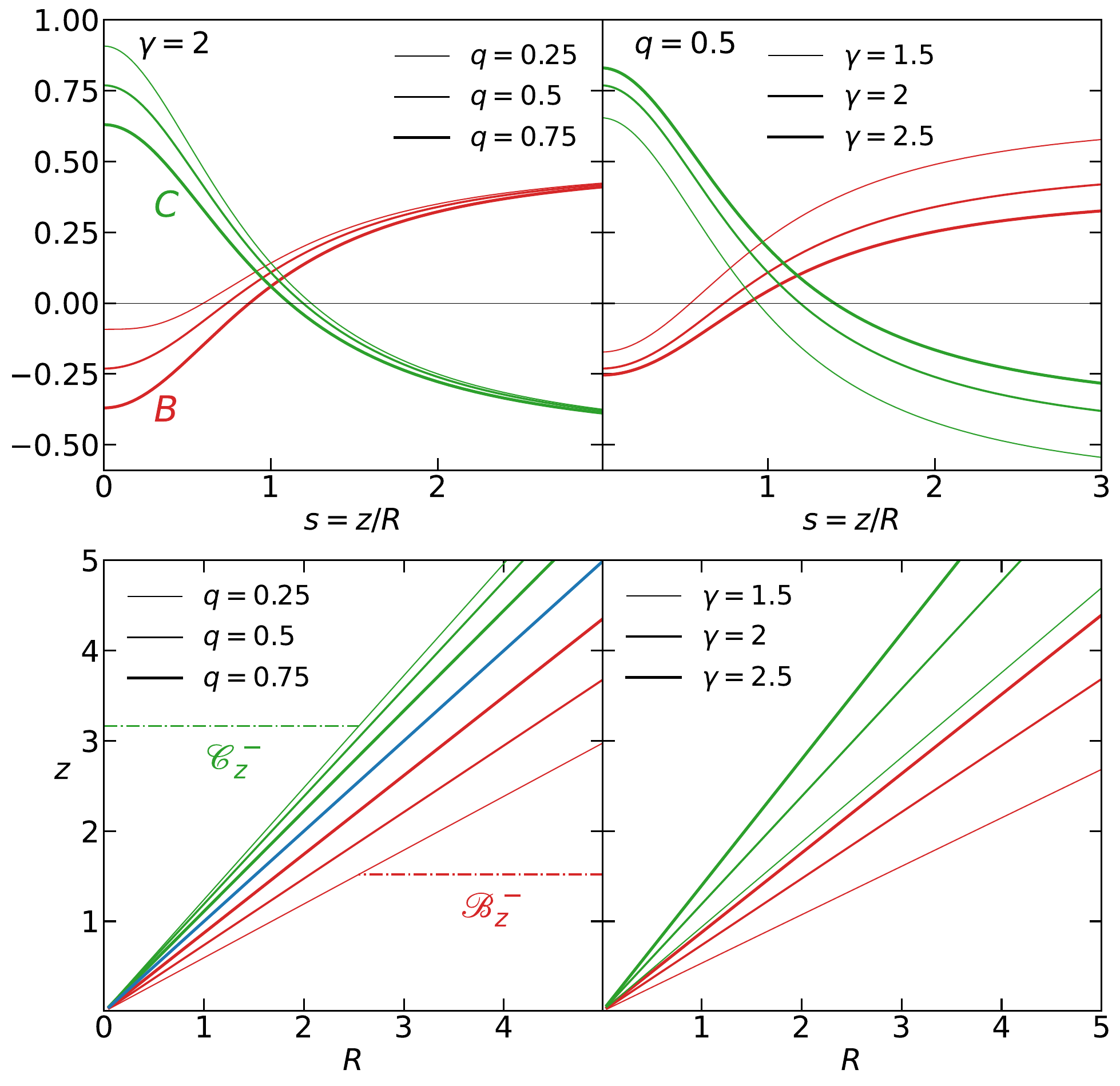}
    \caption{Top panels: the functions $B$ (red) and $C$ (green) for power-law ellipsoidal galaxies of density slope $\gamma$ and minor-to-major axial ratio $q$ in a dominant SIS potential, for $\gamma=2$ and different $q$ (left), and for $q=0.5$ and different $\gamma$ (right). The intersection point of red and green lines with the axis of abscissae gives respectively the critical slope $\sB(q,\gamma)$ and $\sC(q,\gamma)$ of the lines separating the $\bpm$ and $\cpm$ regions. Bottom panels: the $\bzero$ (red) and $\czero$ (green) lines for the same models in the top panels. $\bminus$ is the region of the plane below $\bzero$, and $\cminus$ is the region above $\czero$. The blue line in the left panel corresponds to the spherical stellar model, when $\bzero=\czero$.}
\label{f:figA1}
\end{figure*}

The constraints on $b(z)$ can also be obtained easily. From the global vanishing of the commutator, $\bz = \bu = \bd = 1$ (see Section 2.1.4), while from equation (\ref{eq:bmz_criterion}) with $\gamma > 1$
\begin{equation}
    \bM = {\gamma\over\gamma - 1},
\end{equation}
independently of $z$ and $q$, i.e., $\bM$ decreases for increasing $\gamma$.

\subsection{Power-law ellipsoidal galaxies in a dominant SIS potential}
\label{sec:PL_SIS}

In this second family the dominant DM halo potential is modeled as a SIS by fixing $q=1$ in equation (\ref{eq:phi_sis}); velocities are again normalized to $\vh$. From Exercise 13.31 in C21
\begin{equation}
\begin{dcases}
\rhos\sigz^2 ={q^{\gamma-1} R^{-\gamma}\over 2 (1-q^2)^{\gamma/2}} 
\Bx\left({\gamma\over 2},1-{\gamma\over 2}\right),\\
\Commh = {\gamma q^{\gamma-1} R^{-\gamma -1}\over 2 (1-q^2)^{\gamma/2}}\Bx\left(1+{\gamma\over 2},-{\gamma\over 2}\right),
\end{dcases}
\label{eq:2int_PL_sol}
\end{equation}
where $\Bx(a,b)=\int_0^xt^{a-1}(1-t)^{b-1}dt$ is the Euler incomplete beta function, and 
\begin{equation}\label{eq:x_def}
    x \equiv {(1-q^2)R^2\over R^2 + z^2}=
    {1-q^2\over 1+s^2};\quad s\equiv{z\over R};
\end{equation}
in the spherical limit $\sigz^2 = 1/\gamma$ and $\Commh =0$, while in the oblate case $\Commh\ge 0$ everywhere. From direct evaluation of equations (\ref{eq:d_def}), (\ref{eq:b_def}) and (\ref{eq:c_def}), or from recursive properties of $\Bx(a,b)$, we then obtain
\begin{equation}
D={q^{\gamma-1}\over R^{\gamma +1}}
\left[{s^2(q^2+s^2)^{-\gamma/2}\over 1+s^2}-
{\gamma\Bx\left(\gamma/2,1-\gamma/2\right)\over 2(1-q^2)^{\gamma/2}}
\right],
\label{AppPLS_D}
\end{equation}
\begin{equation}
B={s^2\over 1+s^2}+
{1-\gamma\over 2}
\left({q^2+s^2\over 1-q^2}\right)^{\gamma/2}
\Bx \left({\gamma\over 2},1-{\gamma\over 2}\right),
\label{AppPLS_B}
\end{equation}
\begin{equation}
C={1\over 1+s^2}-{1\over 2}\left({q^2+s^2\over 1-q^2}\right)^{\gamma/2}
\Bx\left({\gamma\over 2},1-{\gamma\over 2}\right),
\label{AppPLS_C}
\end{equation}
so that also for these models the $B$ and $C$ functions are constant on the straight lines $z = s R$, and again $B\geq 0$ for $\gamma\leq 1$. Finally, in the spherical limit of the stellar density, the $D$, $B$, and $C$ functions are given by equations (\ref{eq:pllog_D}) and (\ref{eq:pllog_BC}) with $q=1$.

In the top panels of Figure \ref{f:figA1} the red and green lines show respectively the $B$ and $C$ functions as a function of the slope $s=z/R$ for three different values of $q$ at fixed $\gamma$ (left panel), and for three different values of $\gamma$ at fixed $q$ (right panel). For each model $\sB(q,\gamma)$ and $\sC(q,\gamma)$ are the intersection points of the curves with the abscissae, so that $B\leq 0$ for $s\leq\sB$, and $C\leq 0$ for $s\geq\sC$. Accordingly, in the bottom panels, $\bminus$ is the part of the plane below the red straight line $z=\sB R$, i.e., $s\leq\sB$. Therefore, $\bminus$ expands (i.e., $\sB$ in the top panels moves to the right) both for increasing $q$ and $\gamma$. Similarly, $\cminus$ is the part of the plane above the green lines lines $z=\sC R$, i.e. $s\geq\sC$, so that $\cminus$ expands ($\sC$ in the top panels moves to the left) for increasing $q$, and for decreasing $\gamma$. Notice how the separation between $\bzero$ and $\czero$ (the region $\bplus\cap\cplus$) reduces as $q$ increases, with $\sC=\sB=\sqrt{\gamma-1}$ and $\bzero = \czero$ in the spherical case (blue line).

The constraints on $b(z)$ are easily determined thanks to the fact that all the relevant functions depend on $s=z/R$ only, so that the problem reduces to the study of functions of one variable, all the radial sections such as $\bzminus$ can be expressed in terms of $s$ (e.g. see the bottom panels in Figure \ref{f:figA1}), and the resulting critical values of $b$ are independent of $z$. For example, in the case of equation (\ref{eq:bmz_criterion}), for $\gamma>1$ the set $\bzminus$ translates to the range $0<s<\sB(\gamma,q)$, and the minimum is reached on the equatorial plane $s=0$ with
\begin{equation}
\bM ={2\,(1/q^2-1)^{\gamma/2}\over\mathrm{B}_{1-q^2} \left(\gamma/2,1-\gamma/2\right)},
\end{equation}
and $\bM(1)=\gamma/(\gamma-1)$ in the spherical limit; $\bM$ decreases for increasing $q$ and decreasing $\gamma$. For $\bz$ in equation (\ref{eq:b0z_criterion}) the set $\dzminus$ translates to $0\leq s<\infty$. The minimum is reached for $s\to\infty$, and with some careful expansion we obtain
\begin{equation}
\bz = {2+\gamma\over 2 + \gamma q^2},
\end{equation}
a decreasing function for increasing $q$ and decreasing $\gamma$, with $\bz(1)=1$. Finally, for equations (\ref{eq:b1_def})-(\ref{eq:b2_def}) the intervals are $\sC<s<\infty$ and $0<s<\sB$, respectively, with $\bu = 1$ independently of $\gamma$, and
\begin{equation}
\bd ={1\over\gamma -1}\left[
{2(1/q^2-1)^{\gamma/2}\over\mathrm{B}_{1-q^2} \left(\gamma/2,1-\gamma/2\right)}-1\right],
\end{equation}
for $\gamma >1$, with the minimum reached for $s=0$. As illustration, in the bottom right panel of Figure \ref{f:fig2}, the heavy lines show the critical $b$ values as a function of $q$ for the value $\gamma =2$.

\subsection{Power-law ellipsoidal galaxies in a dominant monopole potential}
\label{sec:pl_mon}

In this third family of models, the dominant DM halo potential is described generically by a monopole potential $\phih=-G M/r$, so that the results can be interpreted or as the case of a central dominant BH, or as the potential at large radii of a finite mass system; at variance with the two previous models, now velocities are normalized to $\sqrt{G M/\as}$. From Exercise 13.31 in C21 we have
\begin{equation}
\begin{dcases}
\rhos\sigz^2 ={q^{\gamma-1} R^{-\gamma-1}\over 2 (1-q^2)^{(\gamma+1)/2}} 
\Bx\left({\gamma +1\over 2},1-{\gamma\over 2}\right),\\
\Commh = {\gamma q^{\gamma-1} R^{-\gamma-2}\over 2 (1-q^2)^{\gamma/2}}\Bx\left({\gamma+3\over 2},-{\gamma\over 2}\right),
\end{dcases}
\label{eq:2int_BH_sol}
\end{equation}
where $x$ is given again in equation (\ref{eq:x_def}).
In the spherical limit $\sigz^2 =1/[(\gamma+1)r]$ and $\Commh =0$.
The $D$, $B$, and $C$ functions are now given by
\begin{equation}\label{AppPLBH_D}
D={q^{\gamma-1}\over R^{\gamma+2}}
\left[{s^2(q^2+s^2)^{-\gamma/2}\over (1+s^2)^{3/2}}-
{\gamma+1\over 2(1-q^2)^{{\gamma+1\over 2}}}\Bx \left({\gamma+1\over 2},1-{\gamma\over 2}\right)
\right],
\end{equation}
\begin{equation}
B={1\over R}
\left[{s^2\over (1+s^2)^{3/2}}-
{\gamma (q^2+s^2)^{\gamma/2}\over 2(1-q^2)^{{\gamma+1\over 2}}}\Bx \left({\gamma+1\over 2},1-{\gamma\over 2}\right)\right],
\label{AppPLBH_B}
\end{equation}
\begin{equation}
C={1\over R}
\left[{1\over (1+s^2)^{3/2}}-
{(q^2+s^2)^{\gamma/2}\over 2(1-q^2)^{{\gamma+1\over 2}}}\Bx \left({\gamma+1\over 2},1-{\gamma\over 2}\right)\right].
\label{AppPLBH_C}
\end{equation}
In the spherical limit the $D$, $B$, and $C$ functions are elementary, and in particular $B$ is negative for $s<\sqrt{\gamma}$, a slope steeper than that in equation (\ref{eq:pl_log_bminus}). We finally repeat the analysis of Appendix \ref{sec:PL_SIS}, confirming the trends of the critical $b$ values with the model axial ratio $q$. 
In particular, 
\begin{equation}
\bM ={2q\,(1/q^2-1)^{(\gamma+1)/2}\over\gamma\mathrm{B}_{1-q^2} \left(1/2+\gamma/2,1-\gamma/2\right)},
\end{equation}
with $\bM(1)=\gamma/(\gamma-1)$. Moreover,
\begin{equation}
\bz = {3+\gamma\over 3 +\gamma q^2}.
\end{equation}
Finally
\begin{equation}
\bd ={1\over\gamma}\left[
{2q(1/q^2-1)^{(1+\gamma)/2}\over\mathrm{B}_{1-q^2} \left(1/2+\gamma/2,1-\gamma/2\right)}-1\right],
\end{equation}
with $\bd(1)=1$. $\bM$, $\bz$, and $\bd$ all decrease for increasing $q$, as can be seen in the bottom right panel of Figure \ref{f:fig2}, for the illustrative case of the stellar spheroid with $\gamma=2$.

\end{document}